\documentclass[conference]{IEEEtran}
\usepackage[
    a4paper,
    left=0.75in,
    right=0.75in,
    top=0.75in,
    bottom=1in
]{geometry}

\usepackage{lipsum}
\usepackage{amsmath,amssymb,amsfonts}
\usepackage{graphicx}
\usepackage[most]{tcolorbox}
\usepackage{lmodern}
\usepackage{algorithm}
\usepackage{algpseudocode}
\usepackage{booktabs}
\usepackage{multirow}
\usepackage{array}
\usepackage{subcaption}
\usepackage[numbers]{natbib}
\usepackage{hyperref}
\usepackage{booktabs} 
\usepackage{multirow}

\newtcolorbox{rqreply}[1]{
    colback=cyan!10,
    colframe=cyan,
    boxrule=0.8pt,
    arc=4pt,
    left=6pt,
    right=6pt,
    top=6pt,
    bottom=6pt,
    fonttitle=\bfseries,
    title={Reply to #1}
}
\usepackage{pifont}
\newcommand{\cmark}{\ding{51}}
\newcommand{\xmark}{\ding{55}}

\begin{document}

\title{Parameters efficient language model with latent cross-attention for IoT intrusion detection}

\author{%
  \IEEEauthorblockN{%
    Frédéric Adjewa\IEEEauthorrefmark{1}\IEEEauthorrefmark{2},
    Moez Esseghir\IEEEauthorrefmark{1}\IEEEauthorrefmark{3},
    Leïla Merghem-Boulahia\IEEEauthorrefmark{1}\IEEEauthorrefmark{3},
    Cheikh Kacfah\IEEEauthorrefmark{2}%
  }%
  \IEEEauthorblockA{%
    \IEEEauthorrefmark{1}University of Technology of Troyes, LIST3N, France\\
    \IEEEauthorrefmark{2}Computer Science Department, Institut Universitaire Saint Jean du Cameroun, Cameroon\\
    \IEEEauthorrefmark{3}3 European Research Center for Cybersecurity and Resilience SECURE+,\\
    European University of Technology, European Union\\
    Email: \{frederic.adjewa, moez.esseghir, leila.merghem\_boulahia\}@utt.fr,\\
    cheikh.kacfah@saintjeaningenieur.org%
  }%
}
\maketitle

\begin{abstract}
Language models have shown strong potential for intrusion detection, but their high memory and computational demands make them difficult to deploy in resource-constrained IoT environments. In this work, we propose a compact Transformer-based architecture that compresses the input into a small latent space before processing it with a minimal Transformer core. We analyze nine compression configurations, compare parameter sharing with layer stacking, and evaluate our approach against SecurityBERT on two IoT intrusion detection datasets, Edge-IIoTset and CICIoT2023. With only $306$k non-embedding parameters and $1.6$\,M parameters in total ($6$\,MB), our model is about $7\times$ smaller than SecurityBERT ($11.2$\,M parameters, $45$\,MB). With the same privacy-preserving encoding, it outperforms SecurityBERT by about $2$ points of macro F1-score on Edge-IIoTset and is within $0.7$ points on CICIoT2023. It is also about $89\times$ faster on a simulated edge device. We further show that a single Transformer layer is sufficient, and that parameter sharing matches layer stacking with up to $66\%$ fewer parameters. These findings show that lightweight Transformer-based architectures can provide a practical and accurate solution for intrusion detection in resource-constrained IoT settings.\footnote{Preprint. Under preparation for journal submission.}.
\end{abstract}


\section{Introduction}

Intrusion detection is no longer an optional component of modern digital infrastructures; it has become a critical requirement. This is especially true in the context of the Internet of Things (IoT), whose rapid expansion is reshaping many sectors of the society \cite{achbarou2025enhanced} As IoT ecosystems continue to grow, securing them becomes increasingly important, since these environments support massive volumes of heterogeneous interactions between connected devices. At the same time, the attack surface expands accordingly, and the threat landscape evolves at a comparable, if not faster, pace. In this context, intrusion detection systems (IDSs) play a central role in protecting the confidentiality, integrity, and availability of networks. They often constitute the first line of defense in IoT and Industrial IoT (IIoT) environments by identifying, reporting, and sometimes reacting to malicious or abnormal activities observed in network traffic \cite{altunay2023hybrid}.
A wide range of intrusion detection approaches has been proposed in the literature, including anomaly-based intrusion detection systems, which aim to identify deviations from normal network behavior. These approaches have benefited considerably from advances in machine learning and deep learning \cite{ajayan2025distilxids}. More recently, language-model-based methods have emerged as a promising direction for improving intrusion detection, since network logs, packets, and related telemetry can often be represented as text-like sequences. However, despite their strong modeling capacity, large language models remain difficult to deploy in IoT environments because of their substantial computational and memory requirements. Several studies have therefore explored ways to reduce this burden, for example, by reducing the number of Transformer blocks \cite{ferrag2024revolutionizing,adjewa2024efficient} or by offloading heavy processing to more resourceful edge or cloud devices \cite{adjewa2026edge,diaf2024beyond}. 

While these strategies are effective to some extent, they raise an important question: is it really necessary to rely on large pretrained language models originally optimized for text data when dealing with network traffic? More specifically, can a compact Transformer-inspired model be designed to learn directly from network data, without requiring pretrained weights, while still achieving competitive intrusion detection performance?

To address this question, we propose an architecture tailored to resource-constrained environments\footnote{https://github.com/freddyAdjh/pelm-iot-ids}. Our model first compresses the input sequence through a cross-attention mechanism into a learnable latent space, allowing the retention of most of the relevant information while reducing the influence of less useful input components. The resulting compressed representation is then processed by a Transformer module operating in the latent space, implemented either as a stack of Transformer blocks or through recursive parameters sharing. This design yields a more compact and memory-efficient architecture while maintaining strong detection performance. To evaluate the effectiveness of the proposed approach, we address the following research questions: 
\begin{itemize}
    \item RQ1: Which compression configuration provides the best trade-off between performance and efficiency?

    \item RQ2: What are the trade-offs between parameters sharing and Transformer stacking?
    \item RQ3: Can a compact model match the detection performance of a larger Transformer-based IDS while requiring substantially fewer computational resources? 
   
\end{itemize}

The main contributions of this paper are threefold. 
\begin{enumerate}
    \item We introduce a compact cross-attention-based architecture for intrusion detection that avoids reliance on large pretrained language models and is specifically designed for constrained IoT deployment;
    \item We provide a systematic analysis of the effects of latent-space size, Transformer depth, and parameters-sharing strategies on model performance, complexity, and runtime efficiency;
    \item We show experimentally that the proposed architecture achieves competitive performance relative to a much larger BERT-based baseline while requiring substantially fewer parameters and offering faster inference, particularly in IoT settings
\end{enumerate}

The remainder of this paper is organized as follows. Section \ref{sec:related_work} reviews related work. Section \ref{sec:methodology} presents the proposed architecture and methodology. Section \ref{sec:experimental_setup} describes the experimental setup. Section \ref{sec:results_discussion} reports and discusses the results with respect to the research questions. Finally, Section \ref{sec:conclusion} concludes the paper and outlines future directions.

\section{Related Work}\label{sec:related_work}

Since their emergence, language models have become a strong foundation for a wide range of text-related security tasks, including intrusion detection, log analysis and threat pattern identification.

Several recent studies have explored transformer-based architectures for intrusion detection. In \cite{ajayan2025distilxids} the authors proposed DistilXIDS, a DistilBERT-based intrusion detection system fine-tuned for Distributed Denial-of-Service (DDoS) attack detection. Their model achieved strong accuracy, recall and precision, but still relied on 66.96 million trainable parameters. Similarly, \cite{wang2026generalizable} introduced a BERT-GNN framework designed to capture both temporal and semantic information. Their approach uses the pretrained bert-base-uncased model, which contains 110 million trainable parameters and experimental results demonstrated strong predictive performance. In \cite{hussain2025ransomware} the authors combined RoBERTa and BERT to detect and classify ransomware in IoT devices. Although the framework produced promising results, it required fine-tuning two large pretrained models, which considerably increases computational cost.

Other works have proposed transformer-inspired or hybrid architectures tailored to IoT and IIoT intrusion detection. In \cite{wang2022microcontroller} the authors addressed intrusion detection in industrial IoT using a microcontroller chip temperature fingerprinting approach. Their framework combines a CNN-based encoder for pattern extraction with a transformer model for final detection, although the model size was not clearly reported. Focusing on MQTT-enabled IoT networks, \cite{ullah2023tnn} proposed a transformer neural network (TNN) that uses self-attention to model dependencies across sequence segments and prioritize the most informative parts of the input. In \cite{wang2023transids} the authors introduced TransIDS, a transformer-based intrusion detection system designed to extract global temporal features directly from IoT traffic flows without relying heavily on manual feature engineering. Likewise, \cite{chai2023ctsf} proposed CTSF for IIoT intrusion detection, combining CNNs and transformers to capture local and global features, while using a support vector machine as the final classifier.

A number of studies have also explored multi-model architectures to enhance detection capability. In \cite{diaf2024beyond} the authors proposed an intrusion detection framework combining BERT, GPT, an LSTM and a standard machine learning classifier. In this design, GPT predicts the next packet autoregressively, BERT evaluates the suitability of these predictions and the LSTM performs binary classification, with the final goal of improving detection performance. A similar direction was followed by \cite{yan2023multi} which combines fine-tuned GPT, BERT and LSTM modules for malicious packet identification. In \cite{wang2023securing} the authors proposed a transformer-based intrusion detection approach that jointly exploits traffic and telemetry data to learn the behavioral patterns and effects of attacks.

Our previous work in \cite{adjewa2026edge} explored a different strategy by offloading transformer processing to the cloud or edge, while the final intrusion prediction was performed on the device using returned embeddings and a lightweight neural network classifier. In related work, \cite{ali2024empowering} investigated the effectiveness of combining BERT with SMOTE and MLP to improve intrusion detection based on machine learning and deep learning techniques. Along the same line, \cite{ali2025enhancing} proposed a framework specifically tailored to intrusion detection in industrial IoT environments.

Table \ref{tab:related_work_comparison} provides a qualitative comparison between our approach and reviewed sutdies.

\begin{table*}[ht!]
\centering
\caption{Qualitative comparison with representative related work. IoT: explicitly designed for IoT or IIoT settings; Pretrained Language Model: relies on pretrained language models that typically require adaptation or fine-tuning on network traffic data; Non-Hybrid: uses homogene architecture; Deployable: suitable for deployment in resource-constrained environments.}
\label{tab:related_work_comparison}
\footnotesize
\setlength{\tabcolsep}{3.5pt}
\begin{tabular}{lcccc}
\toprule
\textbf{Method} & \textbf{IoT} & \textbf{Pretrained model free} & \textbf{Non-hybrid} & \textbf{Deployable} \\
\midrule
\cite{ajayan2025distilxids}      & \xmark & \xmark & \cmark  & \xmark \\
\midrule
\cite{wang2026generalizable}     & \xmark & \xmark & \xmark  & \xmark \\
\midrule
\cite{hussain2025ransomware}     & \cmark & \xmark & \xmark  & \xmark \\
\midrule
\cite{wang2022microcontroller}   & \cmark & \cmark & \xmark  & \cmark \\
\midrule
\cite{ullah2023tnn}              & \cmark & \cmark & \cmark  & \cmark \\
\midrule
\cite{wang2023transids}          & \cmark & \cmark & \cmark  & \xmark \\
\midrule
\cite{chai2023ctsf}              & \cmark & \cmark & \xmark  & \xmark \\
\midrule
\cite{diaf2024beyond}            & \xmark & \xmark & \xmark  & \xmark \\
\midrule
\cite{yan2023multi}              & \xmark & \xmark & \xmark  & \xmark \\
\midrule
\cite{wang2023securing}          & \xmark & \cmark & \cmark  & \xmark \\
\midrule
\cite{adjewa2026edge}            & \cmark & \xmark & \xmark  & \cmark \\
\midrule
\cite{ali2024empowering}         & \xmark & \xmark & \xmark  & \xmark \\
\midrule
\cite{ali2025enhancing}          & \cmark & \xmark & \xmark  & \xmark \\
\midrule
\textbf{Ours}                    & \cmark & \cmark & \cmark  & \cmark \\
\bottomrule
\end{tabular}
\end{table*}

These existing studies show that transformer-based models can achieve strong intrusion detection performance across IoT, IIoT and broader network security settings. However, many of these methods rely on large pretrained language models or multi-component architectures with substantial computational and memory requirements. This limitation motivates the need for more compact transformer-inspired designs that retain strong detection capability while remaining suitable for resource-constrained IoT deployments.

\section{Methodology}\label{sec:methodology}

For clarity and reproducibility, table \ref{tab:symbols} summarizes the notation used throughout the model description and architecture analysis.

\begin{table}[ht!]
\centering
\caption{Summary of symbols and notation used in the study.}
\label{tab:symbols}
\begin{tabular}{ll}
\hline
\textbf{Symbol} & \textbf{Description} \\
\hline
$N$ & Input sequence length \\
$M$ & Number of latent vectors ($M \ll N$) \\
$|\mathcal{V}|$ & Vocabulary size \\
$E$ & Input embedding dimension \\
$e$ & Latent embedding dimension \\
$h$ & Hidden dimension of the feed-forward network \\
$L$ & Depth of Transformer stack \\
\hline
$x \in \mathbb{R}^{N \times E}$ & Input sequence embeddings \\
$z \in \mathbb{R}^{M \times e}$ & Learnable latent matrix \\
$Q, K, V$ & Query, key and value matrices \\
$A$ & Attention weight matrix \\
\hline
$W_Q$ & Query projection matrix \\
$W_K$ & Key projection matrix \\
$W_V$ & Value projection matrix \\
$W_{QK}$ & Combined query-key projection matrix \\
$W_1, W_2$ & Feed-forward network weight matrices \\
\hline
$\#\theta_{\text{emb}}$ & Number of embedding parameters \\
$\#\theta_{\text{CA}}$ & Number of cross-attention parameters \\
$\#\theta_{\text{LT}}$ & Parameters per latent Transformer block \\
\hline
$\phi(\cdot)$ & Softmax activation function \\
$d_k$ & Key dimensionality \\
\hline
\end{tabular}
\end{table}

\subsection{The transformer models and the computation bottleneck}
The impressive performance of Transformer-based models on sequence processing tasks is largely attributed to a single core component: the \emph{attention mechanism} \cite{11315233,10945307,10820504} and its derivatives \cite{warner2025smarter} as defined in eq. \eqref{eq:attention_mech} self-attention enables each token in an input sequence to attend to every other token in that sequence, thereby dynamically modulating the contribution of contextual elements when forming representations.

Formally, given an input sequence encoded into three matrices namely Queries ($Q$), Keys ($K$) and Values ($V$),  the scaled dot-product is computed as
\begin{equation}\label{eq:attention_mech}
    A = \mathrm{softmax}\!\left(\frac{QK^{\top}}{\sqrt{d_k}}\right)V
\end{equation}
where $d_k$ denotes the dimensionality of the Key matrix. The normalization factor $\sqrt{d_k}$ stabilizes the gradient during training by preventing excessively large dot products. The resulting attention matrix assigns a weight to each pair of tokens, quantifying their relative importance in the representation learning process.

To further enhance representational capacity, this mechanism is typically extended to \emph{multi-head attention} wherein multiple attention operations are performed in parallel on distinct linear projections of $Q$, $K$ and $V$. Each head learns complementary relational patterns and their outputs are concatenated and linearly transformed. This design allows finer-grained modeling of dependencies across different subspaces, significantly improving expressiveness compared to single-head self-attention.

Transformer encoder architectures, such as BERT, have been widely adopted in network traffic analysis and cybersecurity applications, including intrusion detection systems \cite{sattarpour2025ebids,li2025efficient,yang2025bert,adjewa2025unseen}. The original BERT model consists of $12$ Transformer encoder blocks and contains approximately $110$ million parameters. For downstream tasks, it is often unnecessary to utilize the entire depth of the model; instead, a subset of layers may be selected or the architecture may be fine-tuned end-to-end to suit task-specific constraints \cite{10423646}

Crucially, for many applications the pretrained BERT is often misused. What matters the most is the use of Transformers. This observation becomes particularly relevant in the IoT environments, where deployment constraints such as limited computational resources, restricted memory and power budgets must be carefully considered. Directly deploying standard Transformer architectures in such settings is generally infeasible due to their substantial computational demands.

The computational bottleneck arises primarily from the self-attention operation, whose complexity scales quadratically with respect to the input sequence length. Specifically, for an input of length $N$, the computation of $QK^{\top}$ is subject to a cost of $\mathcal{O}(N^2 d_k)$ and this operation is repeated across several Transformer blocks. As noted in \cite{jaegle2021perceiver} this quadratic scaling severely limits the applicability of Transformers to long sequences and resource-constrained environments.

To mitigate this computational overhead, we aim at designing a lightweight language model suitable for IoT deployment. Similarly to \cite{lebonvallet2025perceiver} the  approach consists of compressing the input sequence into a significantly smaller set of learnable latent vectors of size $M$, with $M \ll N$. The attention is then computed between the input and this latent representation, resulting in a complexity that scales linearly with the input size, $\mathcal{O}(NM)$ and quadratically with the latent size $\mathcal{O}(M^2)$. This reformulation enables efficient processing while preserving the expressive power of attention-based models.

\subsection{Proposed architecture}

\begin{figure*}[htp!]
        \centering
        \includegraphics[width=\linewidth]{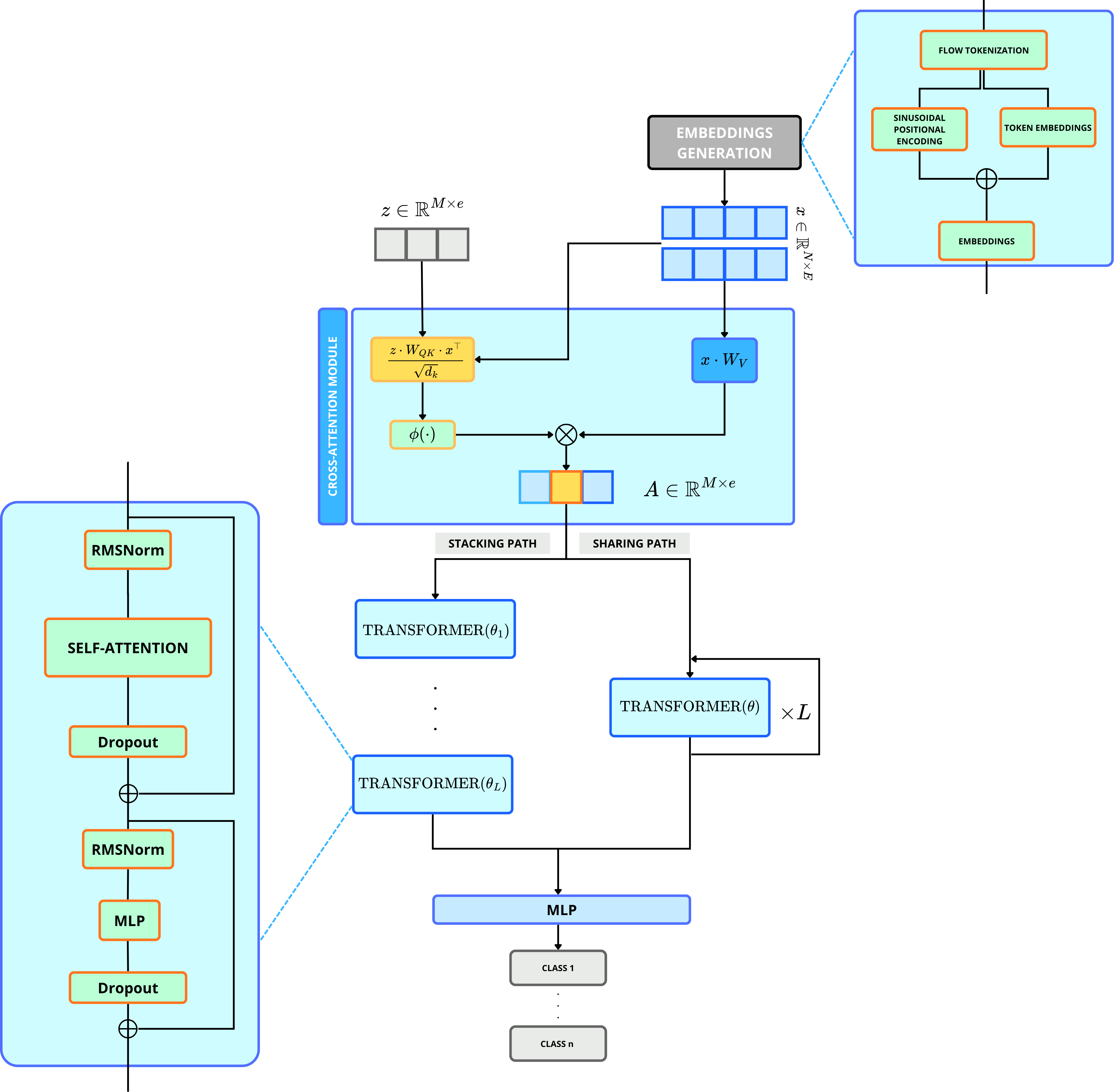}
    \caption{Proposed architecture of the cross-attention-based model for intrusion detection. The framework evaluates two alternative downstream processing paths: a stacking path consisting of a sequential cascade of distinct Transformer blocks, and a sharing path featuring the recurrent use of a single, parameter-shared Transformer block.}
    \label{fig:model_architecture}
\end{figure*}

In our proposed model illustrated on Fig. \ref{fig:model_architecture}, we do not employ multi-head attention. This design choice is motivated by the fact that increasing the number of attention heads does not change the total number of the model parameters. However, we strongly encourage its application because it helps to learn richer representation.

Let the input sequence be denoted by
\[
x \in \mathbb{R}^{N \times E}
\]

where $N$ is the input length and $E$ is the embedding dimension. In the architecture, we also have a set of $M$ learnable latent vectors,
\[
z \in \mathbb{R}^{M \times e}
\]
where $M \ll N$ denotes the number of latent vectors and $e$ is the latent embedding dimension.

The cross-attention module projects the input and latent representations into a feature space of dimension $e$ through three learnable matrices,
\[
W_Q \in \mathbb{R}^{e \times e} \qquad
W_K \in \mathbb{R}^{E \times e} \qquad
W_V \in \mathbb{R}^{E \times e}
\]
Importantly, these matrices do not depend on either the input length $N$ or the latent size $M$, which means that the number of parameters remains independent of the sequence length \cite{lebonvallet2025perceiver}

The query, key and value representations are computed as
\begin{equation}\label{eq:query}
    Q = z W_Q \in \mathbb{R}^{M \times e}
\end{equation}
\begin{equation}\label{eq:key}
    K = x W_K \in \mathbb{R}^{N \times e}
\end{equation}
\begin{equation}\label{eq:value}
    V = x W_V \in \mathbb{R}^{N \times e}
\end{equation}

The context matrix is then obtained using scaled dot-product attention,
\begin{equation}\label{eq:att_scores}
    C = \sigma\!\left(\frac{QK^{\top}}{\sqrt{d_k}}\right) \in \mathbb{R}^{M \times N}
\end{equation}
where $\sigma(\cdot)$ denotes the row-wise softmax function. Since $Q \in \mathbb{R}^{M \times e}$ and $K \in \mathbb{R}^{N \times e}$, the computation of $QK^{\top}$ scales linearly with the input length $N$ for fixed latent size $M$ and latent dimension $e$.

The latent attention is then computed as
\begin{equation}\label{eq:context}
    z_A = C V \in \mathbb{R}^{M \times e}
\end{equation}
This operation produces a compact and enriched representation of the input sequence whose dimensionality depends only on the latent size $M$, thereby decoupling the downstream representation from the original input length $N$.

After this cross-attention stage, the latent representation is processed by a sequence of Transformer blocks operating exclusively in the latent space. At this stage, the model uses standard self-attention, whose computational complexity is quadratic with respect to the latent size, i.e., $\mathcal{O}(M^2)$ and is therefore independent of the original input length $N$.

Two architectural strategies can be considered for this post-cross-attention processing illustrated by the two possible paths on the model architecture figure.

\paragraph{Transformer stacking strategy}
Here, $L$ Transformer blocks are stacked sequentially, with each block having its own learnable parameters. The latent representation is updated as
\[
z_A^{(\ell)} = \mathcal{T}_{\ell}\!\left(z_A^{(\ell-1)}\right), \qquad \ell = 1, \dots, L
\]
where $\mathcal{T}_{\ell}$ denotes the $\ell$-th Transformer block with parameters $\theta_{\ell}$. This design increases representational flexibility, but it also causes the number of parameters to grow linearly with $L$, which increases both model size and memory requirements.

\paragraph{Parameters-sharing strategy}
In the second strategy, the same Transformer block is applied repeatedly $L$ times, so that parameters are shared across all layers. The update rule becomes
\[
z_A^{(\ell)} = \mathcal{T}\!\left(z_A^{(\ell-1)}\right), \qquad \ell = 1, \dots, L
\]
where a single set of parameters $\theta$ is reused throughout the $L$ iterations. This formulation can be interpreted as an iterative refinement of the latent representation through repeated application of the same transformation. A key advantage of this strategy is that the number of parameters remains constant regardless of $L$, making it particularly suitable for resource-constrained deployment settings.

\subsection{Model paramters assessing}

In this section, we analyze the number of learnable parameters of the proposed architecture under the two stacking strategies introduced above. 
We begin by examining the parameterization of the embedding layer, which is common to all sequence-based models.

\subsubsection{Embeddings}

For sequential data, each input element must first be mapped into a continuous vector space through an embedding layer.

\paragraph{Token embeddings}
Let $V$ denote the vocabulary size and $E$ the embedding dimension. The token embedding matrix maps each discrete token identifier to an $E$-dimensional vector. Therefore, the number of parameters associated with token embeddings is
\[
\#\theta_{\text{token}} = E|\mathcal{V}|
\]

This parameters count scales linearly with the vocabulary size and the maximum sequence length. 

\paragraph{Positional Embeddings}
To incorporate ordering information into the input sequence, we encode the relative position of each token using deterministic positional embeddings. Following \cite{vaswani2017attention} we adopt sinusoidal functions to generate position-dependent representations, as defined in eq.( \ref{eq:positional}). In contrast to learned positional encodings, this formulation introduces no additional trainable parameters while achieving comparable empirical performance.

\begin{equation}
\label{eq:positional}
\begin{aligned}
    \mathrm{PE}(pos, 2i) &= \sin\left(pos/{10000^{2i/d_{model}}}\right), \\
    \mathrm{PE}(pos, 2i+1) &= \cos\left(pos/{10000^{2i/d_{model}}}\right)
\end{aligned}
\end{equation}
where $pos$ denotes the position of the token within the sequence, $i$ indexes the embedding dimension and $d_{\mathrm{model}}$ represents the dimensionality of the model embeddings.

\subsubsection{Cross-Attention module}

As introduced previously, the cross-attention module is the key component that enables input compression and alleviates the quadratic scaling issue inherent to standard self-attention mechanisms. By attending from a small set of latent vectors to the full input sequence, this module ensures that the computational and parametric complexity remains independent of the input length.

In our proposed formulation\footnote{Bias terms are deliberately omitted in the parameters analysis, as they contribute negligibly to the total parameters magnitude.} The following components contribute as follow:

\paragraph{Value projection}
The value projection maps the input embeddings to the latent feature space, resulting in
\begin{equation}
\#\theta_V = eE
\end{equation}
learnable parameters.

\paragraph{Query and key projections}
Naively, the query and key projections would require two separate matrices,
\(
W_Q, 
W_K 
\)
leading to a total of $e^2 + E e$ parameters. However, cross-attention scores are computed as
\begin{equation}
    QK^{\top} = (z W_Q)(x W_K)^{\top}
\end{equation}
By rearranging the terms, this expression can be rewritten as
\begin{equation}\label{eq:trick}
    QK^{\top}
    = z W_Q W_K^{\top} x^{\top}
    = z W_{QK} x^{\top}
\end{equation}
where
\begin{equation}
W_{QK} = W_Q W_K^{\top} \in \mathbb{R}^{e \times E}
\end{equation}
This reformulation allows the query–key interaction to be parameterized by a single projection matrix $W_{QK}$, instead of two separate matrices. Consequently, the number of learnable parameters associated with the attention score computation becomes
\begin{equation}
\#\theta_{QK} = eE
\end{equation}

\paragraph{Total cross-attention parameters}
Under this parameterization, the total number of learnable parameters in the cross-attention module is
\begin{equation}
\#\theta_{\text{CA}} = 2 eE
\end{equation}

\subsubsection{Latent transformer module}

The latent Transformer \footnote{Latent because it processes latent component as input.} module follows the standard Transformer encoder design, restricted to a single self-attention head and operates exclusively on the enriched latent representation produced by the cross-attention module.

Let $z^{(0)}_A \in \mathbb{R}^{M \times e}$ denote the output after cross-attention. Each latent Transformer block consists of a vanilla self-attention layer followed by a multi layer perceptron layer (MLP).
\footnote{Residual connections and normalization layers are omitted from the parameters count, as their contribution is negligible compared to the dominant linear projections.}

\paragraph{Self-attention parameters}
In the single-head self-attention setting the total number of parameters associated with latent self-attention is 
\(
\#\theta_{\text{SA}} = 3e^2
\)

Following this module, the self-Output block has $e^2$ number of parameters.

\paragraph{Multi-layer perceptron parameters}
The multi layer perceptron layer consists of two linear transformations with an intermediate hidden dimension $h$, typically chosen as a multiple of $e$. In our setup, similar to \cite{vaswani2017attention} we use $h=4e$. 
Resulting in
\(
\#\theta_{\text{MLP}} = 8 e^2
\)
learnable parameters.

\paragraph{Parameters per latent Transformer block}
Combining the self-attention and feed-forward components, the total number of parameters for a single latent Transformer block is
\( 
\#\theta_{\text{LT}} = 12e^2
\)
Following the architectural choices discussed earlier,
\begin{itemize}
    \item Stacking strategy: stacking $L$ latent Transformer blocks with distinct parameters sets leads to a total of
    \begin{equation}
        \#\theta_{\text{LT}}^{\text{indep}} = 12L e^2
    \end{equation}
    
    parameters.
    \item Sharing strategy: reusing a single latent Transformer block across $L$ iterations yields
     \begin{equation}
    \#\theta_{\text{LT}}^{\text{shared}} = 12 e^2
    \end{equation}
    making the parameters count independent of $L$.
\end{itemize}

This analysis highlights that, under parameters sharing, increasing the depth of latent processing may improve representational refinement without increasing the model size, a property that is particularly advantageous for our purpose which is the deployment in resource-constrained IoT environments. 

In summary, the rough parameter count of the model is given by (\ref{eq:param_count}):
\begin{equation}\label{eq:param_count}
     \#\theta_{total} =  E|\mathcal{V}| + 2eE + 12e^2 + \Delta(e)
\end{equation}
where $\Delta(e)$ represents the residual parameters omitted from our evaluation (bias, residual connections and normalization layers).




Finally, when adopting the weight-sharing strategy, the parameters contributions of both the attention and cross-attention modules decrease substantially. In this configuration, the embedding layers clearly become the main parameters bottleneck.


\subsection{Other internal improvements}
In the standard transformer encoder architectures, to stabilize the learning and improve the convergence, Layer normalization have been used and have shown good performance. However, the re-scaling and the re-centering introduce computational overhead \cite{zhang2019root} in our model architecture, all the normalization have been replaced by the RMSNorm which only focuses on the re-scaling, hence improves the running time.

\section{Experimental design}\label{sec:experimental_setup}

To evaluate our model we leverage two widely used IoT intrusion-detection benchmarks.

The CICIoT2023 dataset \cite{neto2023ciciot2023} is a large-scale benchmark collected from a realistic experimental testbed. It contains approximately 33 different attack types spanning multiple communication protocols and includes more than 46 million records. Due to its size, we construct a representative subset by performing stratified random sampling across attack categories to preserve the original class distribution.

The EdgeIIoTset dataset \cite{ferrag2022edge} is another benchmark specifically designed for IoT intrusion detection. It contains 14 attack types in addition to benign traffic. Together, CICIoT2023 and EdgeIIoTset provide a suitable evaluation setting for measuring the effectiveness of our approach in IoT environments.



\subsection{Data preparation}

For all datasets, we retain only attack classes with at least $1000$ samples to ensure sufficient representation during training. Each dataset is split into training and testing sets using a $70:30$ ratio. Since benign/normal traffic accounts for the majority of entries in intrusion detection datasets, we deliberately retain only a fraction of that class to focus primarily on detecting malicious traffic.

Because our approach relies on transformer-based models, we construct a vocabulary for each dataset. Following \cite{adjewa2025unseen,ferrag2024revolutionizing,adjewa2024efficient} we limit the vocabulary size to 5000 tokens in order to maintain a compact and computationally efficient representation. Moreover, the algorithm used to prepare the data for ingestion by the model uses a word-level tokenizer trained on 70\% ratio of the encoded the data.

\subsection{Experimental configuration and evaluation metrics}

To assess model performance with respect to compression, we vary the two main compression-related hyperparameters, $e$ and $M$, over the set $\{64, 128, 256\}$.

Our experiments were conducted for training and part of the evaluation on an NVIDIA GH100 GPU with 120 GB of memory. To provide a more realistic assessment of deployment feasibility, we also evaluated inference on a 4-core Raspberry Pi 3 with 1 GB of memory. In addition, we simulated an edge device equipped with 512 MB of memory and a single-core CPU. The model was implemented in PyTorch~2.6 with CUDA~12 support. The detailed experimental configuration is reported in Table \ref{tab:model_config}. Inference was performed using ONNX Runtime 1.19.2.

In each experiment, the model was trained for a certain number of epochs depending on the goal or until it converges. Early stopping was applied when the convergence criterion was met in order to prevent overfitting. In addition, a learning rate scheduler was used to periodically reduce the learning rate during training, which stabilizes optimization and helps avoid convergence plateaus.

To evaluate model performance, we mainly rely on the macro-averaged F1-score defined in eq.~(\ref{eq:mF1}). Because the datasets used in this work are class-imbalanced, accuracy alone can provide a misleading estimate of classification efficacy. The macro-F1 score assigns equal weight to all classes, effectively capturing both precision and recall across all attack categories. 

In addition to macro-F1, we evaluate the False Positive Rate (FPR), False Negative Rate (FNR), and Detection Rate (DR) to comprehensively assess detection capabilities across individual classes:

\begin{equation}
    \text{DR}_i = \frac{\text{TP}_i}{\text{TP}_i + \text{FN}_i} 
\end{equation}\label{eq:dr}
\begin{equation}\label{eq:fpr} 
    \text{FPR}_i = \frac{\text{FP}_i}{\text{FP}_i + \text{TN}_i}
\end{equation}
\begin{equation}
    \text{FNR}_i = 1 - \text{DR}_i
\end{equation}\label{eq:fnr}

where $\text{TP}_i$, $\text{TN}_i$, $\text{FP}_i$, and $\text{FN}_i$ denote the true positives, true negatives, false positives, and false negatives for class $i$, respectively.

Finally, the overall macro-F1 score across all $N$ classes is given by:
\begin{equation}\label{eq:mF1}
    \text{Macro F1} = \frac{1}{N} \sum_{i=1}^{N} \text{F1}_i
\end{equation}

with,

\begin{equation}
    \text{F1}_i = \frac{2 \cdot \text{Precision}_i \cdot \text{DR}_i}{\text{Precision}_i + \text{DR}_i}
\end{equation}

To assess computational efficiency, we report the number of floating-point operations (FLOPs), which provides a hardware-independent measure of model complexity.

\begin{table}[ht!]
  \centering
  \small
  \setlength{\tabcolsep}{4pt}
  \caption{Model and training configuration (cross-attention with weight sharing).}
  \label{tab:model_config}
  \begin{tabular}{@{}ll@{}}
    \toprule
    \textbf{Parameter} & \textbf{Value} \\
    \midrule
    \multicolumn{2}{@{}l}{\textit{Architecture}} \\
    Input sequence length ($N$)      & 512 \\
    Input embedding dim.\ ($E$)      & 256 \\
    Latent embedding dim.\ ($e$)     & $\{64, 128, 256\}$ \\
    Number of latent vectors ($M$)   & $\{64, 128, 256\}$ \\
    Transformer layers ($L$)         & $\{1, 2, 4\}$ \\
    Attention heads                  & 1 \\
    Feed-forward dim.                & $4e$ \\
    Dropout                          & 0.1 \\
    \midrule
    \multicolumn{2}{@{}l}{\textit{Training}} \\
    Optimizer                        & AdamW \\
    Learning rate                    & $3 \times 10^{-5}$ \\
    LR decay factor                  & 0.1 (on val.\ F1 plateau) \\
    Batch size                       & 32 \\
    Epochs                           & 10 \\
    Early stopping                   & Yes \\
    \bottomrule
  \end{tabular}
\end{table}

\section{Experimental Results}\label{sec:results_discussion}

\subsection{Which compression configuration provides the best trade-off between performance and efficiency?}

To answer this question, we evaluated several architectural configurations defined by the latent embedding size ($e$) and the number of latent vectors ($M$), with $e, M \in \{64, 128, 256\}$, which gives nine configurations in total. In all cases, the minimal architecture consists of a cross-attention module followed by a single transformer block. All experiments were conducted on the Edge-IIoTset dataset. Each configuration was trained for 10 epochs with three random seeds (0, 1, 2), and we report the mean $\pm$ standard deviation over the three runs. We deliberately limited training to 10 epochs to assess how quickly each configuration converges.

\begin{figure}[ht!]
    \centering
    \begin{subfigure}{0.48\columnwidth}
        \centering
        \includegraphics[width=\linewidth]{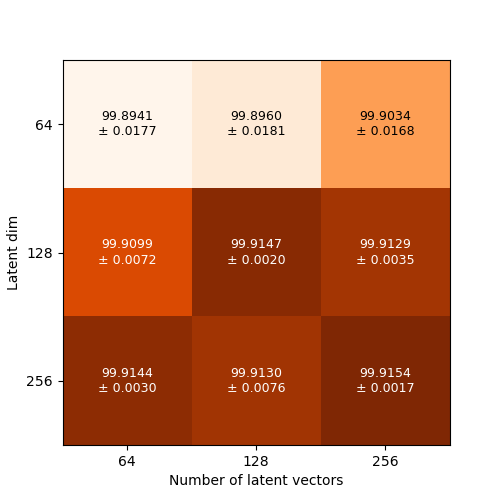}
        \caption{Accuracy}
        \label{fig:accuracy}
    \end{subfigure}
    \hfill
    \begin{subfigure}{0.48\columnwidth}
        \centering
        \includegraphics[width=\linewidth]{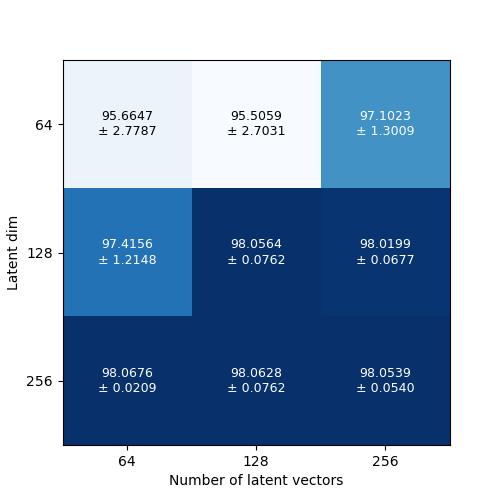}
        \caption{Macro F1-score}
        \label{fig:f1_score}
    \end{subfigure}
    \caption{Accuracy and macro F1-score (\%) of the proposed method for each combination of latent embedding size ($e$) and number of latent vectors ($M$)}
    \label{fig:performance}
\end{figure}

Fig.~\ref{fig:performance} summarizes the performance of all configurations in terms of accuracy and macro F1-score. As shown in Fig.~\ref{fig:accuracy}, all configurations reach an accuracy above $98\%$, even the smallest ones. However, accuracy can be misleading here: the dataset is highly imbalanced, so this metric is barely affected by errors on minority classes. The macro F1-score weights all classes equally and therefore gives a more informative assessment of detection quality. With this metric, the differences between configurations become more visible. On Fig.~\ref{fig:f1_score}, all configurations still reach a macro F1-score above $95\%$. The $e128M128$ configuration reaches $98.06 \pm 0.08\%$, whereas the most compact configuration, $e64M64$, reaches $95.66 \pm 2.78\%$.

As shown in the previous sections, the embedding layer accounts for a large share of the compressed model. Its size, $E|\mathcal{V}|$ (about $1.28$ million parameters), is the same for all configurations. We therefore exclude it when comparing model sizes and refer to the remaining parameters as core parameters.

\begin{figure}[ht!]
    \centering
    \includegraphics[width=\linewidth]{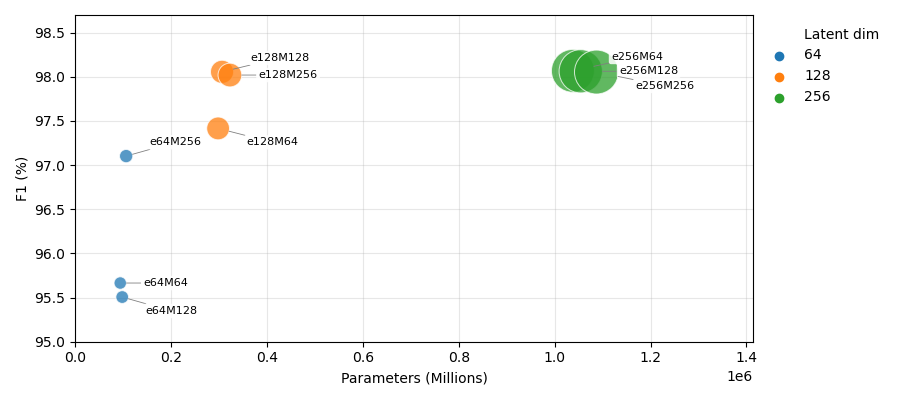}
    \caption{Macro F1-score as a function of the number of core parameters (embedding layer excluded) for each configuration.}
    \label{fig:sizes}
\end{figure}

Fig.~\ref{fig:sizes} shows the macro F1-score as a function of the number of core parameters. The configurations form three distinct clusters, one per latent embedding size. The blue cluster, with the smallest bubbles, contains the most compact configurations ($e=64$), whereas the configurations with $e=256$ form the largest cluster.

\begin{table}[ht!]
  \centering
  \caption{Model size (embedding layer included) and inference compute for each compression configuration.}
  \label{tab:model_cost}
  \begin{tabular}{lrr}
    \toprule
    Configuration & Model size (MB) & GFLOPs / sample \\
    \midrule
    $e64M64$   & 5.24 & 0.046 \\
    $e64M128$  & 5.26 & 0.077 \\
    $e64M256$  & 5.29 & 0.145 \\
    \midrule
    $e128M64$  & 6.02 & 0.088 \\
    $e128M128$ & 6.05 & 0.147 \\
    $e128M256$ & 6.11 & 0.277 \\
    \midrule
    $e256M64$  & 8.84 & 0.214 \\
    $e256M128$ & 8.91 & 0.369 \\
    $e256M256$ & 9.03 & 0.705 \\
    \bottomrule
  \end{tabular}
\end{table}

Table~\ref{tab:model_cost} reports the storage and compute cost of each configuration. The two costs respond differently to the two compression parameters. Model size is driven almost entirely by the latent embedding size: at a fixed $e$, increasing $M$ from 64 to 256 adds at most $0.19$~MB. Compute, in contrast, grows with both parameters. At a fixed $e$, each doubling of $M$ increases the GFLOPs by a factor of $1.7$ to $1.9$. At a fixed $M$, each doubling of $e$ increases them by a factor of $1.9$ to $2.5$. As a result, configurations with similar compute can differ in size. For example, $e64M256$ and $e128M128$ require almost the same compute ($0.145$ vs.\ $0.147$ GFLOPs per sample), but $e128M128$ is $0.76$~MB larger.

The highest macro F1-score is obtained by $e256M64$, in the orange cluster. However, its gain over $e128M128$ is negligible. In exchange, it requires about $3.4\times$ more core parameters and about $1.5\times$ more compute ($0.214$ vs.\ $0.147$ GFLOPs per sample). At the other end of the range, $e64M64$ requires about $3.3\times$ fewer core parameters and $3.2\times$ less compute than $e128M128$. However, its macro F1-score is $2.4$ percentage points lower and varies much more across seeds. This lower score may partly reflect the 10 epochs training budget, as compact configurations may need more training steps to converge. Overall, $e128M128$ offers the best trade-off between detection performance and computational cost among the evaluated configurations.

\begin{rqreply}{RQ1}
The configuration with a latent embedding size of $e=128$ and $M=128$ latent vectors ($e128M128$) provides the best trade-off between detection performance and efficiency. It reaches a macro F1-score of $98.06 \pm 0.08\%$ with approximately $3.1 \times 10^{5}$ core parameters (excluding $E|\mathcal{V}|$, see eq.~(\ref{eq:param_count})) and $0.147$ GFLOPs per sample. 
The best-performing configuration, $e256M64$, improves the macro F1-score only marginally, at the cost of about $3.4\times$ more core parameters and $1.5\times$ more compute. The most compact configuration, $e64M64$, reduces compute by a factor of $3.2$ but lowers the macro F1-score by $2.4$ percentage points.
\end{rqreply}
\subsection{What are the trade-offs between parameters sharing and Transformer stacking?}

To assess the relative benefits of these two approaches, we evaluated the model $e128M128$ under two paradigms: transformer stacking and parameter sharing (recursive formulation). In both cases, we varied the parameters $L$, which controls either the number of transformer layers in the stacking approach or the number of recurrent applications of the same layer in the sharing approach.

\begin{table}[H]
  \centering
  \small
  \setlength{\tabcolsep}{4.5pt}
  \caption{Evaluation macro F1-score (\%) of the shared and stacked strategies for increasing depth $L$ ($e128M128$).}
  \label{tab:rq2_sharing}
  \begin{tabular}{c c cc cc}
    \toprule
     & & \multicolumn{2}{c}{Edge-IIoTset} & \multicolumn{2}{c}{CICIoT2023} \\
    \cmidrule(lr){3-4} \cmidrule(lr){5-6}
    $L$ & GFLOPs & Shared & Stacked & Shared & Stacked \\
    \midrule
    1 & 0.147 & \multicolumn{2}{c}{98.06} & \multicolumn{2}{c}{93.20} \\
    2 & 0.206 & 98.09 & 98.10 & 93.47 & 93.35 \\
    4 & 0.323 & 97.65 & 97.98 & 93.55 & 93.59 \\
    \bottomrule
  \end{tabular}
\end{table}

Table~\ref{tab:rq2_sharing} compares the two strategies in terms of compute and macro F1-score. Both strategies add the same amount of computation for each additional layer. At $L{=}1$, they are the same model. The inference cost increases from $0.147$~GFLOPs per sample at $L{=}1$ to $0.206$ at $L{=}2$ and $0.323$ at $L{=}4$, i.e., about $0.059$~GFLOPs per additional layer. These values are identical for both strategies at a given $L$. Peak training memory follows the same pattern for both strategies: it rises from 185~MB at $L{=}1$ to about 227~MB at $L{=}2$ and about 320~MB at $L{=}4$.
The two strategies differ only in their number of parameters. With parameter sharing, the same layer is applied recursively, so the model keeps 306k non-embedding parameters regardless of $L$. With stacking, each additional layer adds about 198k parameters. For the same compute, sharing therefore uses $39\%$ fewer non-embedding parameters than stacking at $L{=}2$ and $66\%$ fewer at $L{=}4$.

On Edge-IIoTset, increasing $L$ does not improve the macro F1-score. The single-layer model reaches $98.06\%$. At $L{=}2$, the shared and stacked models reach $98.09\%$ and $98.10\%$, respectively, which is within $0.05$ points of the single-layer model. At $L{=}4$, the macro F1-score decreases to $97.65\%$ (shared) and $97.98\%$ (stacked). However, the training curves at $L{=}4$ show that the models were still improving when the 10 epochs budget ended, so this decrease may reflect insufficient training rather than a limitation of deeper models. Regardless of the strategy and of $L$, the same classes remain difficult to detect: \textit{Fingerprinting}, \textit{Ransomware} and \textit{Backdoor}.

On CICIoT2023, macro F1-scores are lower, ranging from $93.20\%$ to $93.59\%$ across all configurations. Increasing $L$ from 1 to 4 improves the macro F1-score by $0.35$ points with sharing and $0.39$ points with stacking. As on Edge-IIoTset, some classes, such as \textit{Recon-OSScan}, \textit{Recon-PortScan} and \textit{DNS Spoofing}, remain difficult to detect for every value of $L$. This indicates that additional depth does not resolve these classes.

\begin{rqreply}{RQ2}
Parameter sharing matches Transformer stacking at every depth on both datasets. Since both have the same compute cost at a given $L$, sharing is the more efficient choice, requiring 39\% to 66\% fewer non-embedding parameters. Increasing $L$ under either strategy raises compute by up to $2.2\times$ without a clear gain on Edge-IIoT and with at most a small gain (${\approx}0.3$ points) on CICIoT23. A shallow recursion with shared parameters therefore offers the best trade-off within the tested configurations, although these results were obtained under a 10 epochs training budget that most runs reached.
\end{rqreply}

\subsection{Can a compact model match the detection performance of a larger Transformer-based IDS while requiring substantially fewer computational resources?}

This section evaluates whether the proposed architecture can serve as a practical alternative to larger legacy Transformer-based intrusion detection systems. To answer this question, we first position our method against representative baselines reported on Edge-IIoTset, including conventional machine learning models, boosting-based approaches, and Transformer-based IDSs. We then compare our model more directly with SecurityBERT in terms of predictive performance, parameter count, model size, inference latency, and memory consumption across several hardware settings. Thourghout the section the model use remains the $e128M128$. The goal is not only to assess raw detection performance, but also to determine whether any performance gap is meaningful enough to outweigh the efficiency benefits of the reduced model in realistic deployment scenarios.

\begin{table*}[ht!]
\centering
\caption{Positioning of our IDS against representative baselines on Edge-IIoTset. For studies that did not report the F1-score, the reported value corresponds to the average of the available class-wise or per-setting results.}
\label{tab:edge_comparison}
\begin{tabular}{llccc}
\hline
\textbf{Reference} & \textbf{Model} & \textbf{Params} & \textbf{Acc (\%)} & \textbf{F1 (\%)} \\
\hline
\multirow{4}{*}{\cite{ferrag2022edge}} 
& DT  & -    & 67.11   & 55.13 \\
\cline{2-5}
& RF  & -    & 80.83   & 75.07 \\
\cline{2-5}
& DNN & -    & 94.67   & 77.33 \\
\cline{2-5}
& SVM & -    & 77.61   & 77.33 \\
\hline
\multirow{3}{*}{\cite{al2023comparative}} 
& AdaBoost        & -    & 86.29   & 86.3 \\
\cline{2-5}
& LogitBoost      & -    & 85.40   & 85.6 \\
\cline{2-5}
& Attributed selected classifier & -    & 90.43   & 90.6 \\
\hline
\cite{ferrag2024revolutionizing} 
& SecurityBERT (with PPFLE) & 11.2M & 98.2   & 84 \\
\hline
\multirow{2}{*}{\cite{adjewa2024efficient}} 
& Federated SecurityBERT (with PPFLE)         & 11.2M & 97.79   & -  \\
\cline{2-5}
& Federated quantized SecurityBERT (with PPFLE) & 11.2M & 97.77    & -  \\
\hline
\multirow{2}{*}{\textbf{Ours}}
& \textbf{Proposed architecture (with PPFLE)} & \textbf{1.6M} & \textbf{98.88} & \textbf{94.55} \\
\cline{2-5}
& \textbf{Proposed architecture (without PPFLE)} & \textbf{1.6M} & \textbf{99.9} & \textbf{98.06} \\
\hline
\end{tabular}
\end{table*}

Table~\ref{tab:edge_comparison} compares the proposed architecture with the considered baselines on Edge-IIoTset. Classical machine learning models (decision trees, random forests and support vector machines) and standard deep neural networks reach F1-scores below $78\%$. Although these models are lightweight, they appear unable to capture the complexity of modern IoT traffic. The boosting-based methods reported in \cite{al2023comparative} improve on these results, but remain clearly below our approach. This suggests that conventional models, even ensemble-based ones, cannot match approaches that explicitly model the sequential structure of network flows.

In contrast, the Transformer-based approaches reported in \cite{ferrag2024revolutionizing,adjewa2024efficient} reach higher performance, which confirms the value of attention-based sequence modeling for intrusion detection. However, these gains come with a large number of parameters: SecurityBERT-based models require about $11.2$\,M parameters. Our architecture keeps the benefits of Transformer-style modeling with about $1.6$\,M parameters, i.e., about $7\times$ fewer.

Without the privacy-preserving fixed-length encoding (PPFLE), which encodes each network flow into a hashed, fixed-length representation before it is provided to the model, our model achieves the best performance among the compared methods: $99.9\%$ accuracy and $98.06\%$ macro F1-score. With PPFLE, which is also the input encoding used by SecurityBERT~\cite{ferrag2024revolutionizing}, it still reaches $98.88\%$ accuracy and $94.55\%$ macro F1-score. These results indicate that our architecture delivers competitive intrusion detection performance in both the standard and the privacy-preserving settings.

\begin{table*}[ht!]
  \centering
  \small
  \caption{Per-class test results on Edge-IIoT for the SecurityBERT baseline and the proposed model (e128M128, shared, $L{=}1$)}
  \label{tab:rq3_edge_per_class}
  \begin{tabular}{l r cccc cccc}
    \toprule
     & & \multicolumn{4}{c}{SecurityBERT} & \multicolumn{4}{c}{\textbf{Ours}} \\
    \cmidrule(lr){3-6} \cmidrule(lr){7-10}
    Class & Support & F1 & DR & FNR & FPR & F1 & DR & FNR & FPR \\
    \midrule
    Norm & 242{,}347 & 100.00 & 100.00 & 0.00 & 0.000 & 100.00 & 100.00 & 0.00 & 0.000 \\
    UDP & 18{,}235 & 100.00 & 100.00 & 0.00 & 0.000 & 100.00 & 100.00 & 0.00 & 0.000 \\
    ICMP & 17{,}466 & 99.99 & 100.00 & 0.00 & 0.001 & 100.00 & 99.99 & 0.01 & 0.000 \\
    SQL & 7{,}681 & 100.00 & 100.00 & 0.00 & 0.000 & 100.00 & 100.00 & 0.00 & 0.000 \\
    Pass & 7{,}523 & 99.95 & 100.00 & 0.00 & 0.002 & 99.95 & 100.00 & 0.00 & 0.002 \\
    TCP & 7{,}510 & 99.97 & 99.99 & 0.01 & 0.001 & 99.96 & 100.00 & 0.00 & 0.002 \\
    HTTP & 7{,}486 & 99.91 & 99.87 & 0.13 & 0.001 & 99.91 & 99.83 & 0.17 & 0.000 \\
    Vuln & 7{,}517 & 99.95 & 99.92 & 0.08 & 0.001 & 99.92 & 99.84 & 0.16 & 0.000 \\
    Uplo & 5{,}645 & 99.99 & 100.00 & 0.00 & 0.000 & 99.98 & 99.96 & 0.04 & 0.000 \\
    Back & 3{,}729 & 97.29 & 96.86 & 3.14 & 0.026 & 97.38 & 97.18 & 2.82 & 0.027 \\
    Port & 3{,}384 & 99.35 & 99.88 & 0.12 & 0.012 & 99.27 & 99.82 & 0.18 & 0.013 \\
    XSS & 2{,}387 & 99.77 & 99.83 & 0.17 & 0.002 & 99.69 & 100.00 & 0.00 & 0.005 \\
    Rans & 1{,}639 & 93.99 & 95.00 & 5.00 & 0.035 & 93.99 & 94.51 & 5.49 & 0.033 \\
    Fing & 150 & 83.40 & 72.00 & 28.00 & 0.000 & 84.17 & 72.67 & 27.33 & 0.000 \\
    MITM & 182 & 100.00 & 100.00 & 0.00 & 0.000 & 100.00 & 100.00 & 0.00 & 0.000 \\
    \midrule
    Macro average & - & 98.24 & 97.56 & 2.44 & 0.005 & 98.28 & 97.59 & 2.41 & 0.005 \\
    Benign vs.\ attack & - & 100.00 & 100.00 & 0.00 & 0.001 & 100.00 & 100.00 & 0.00 & 0.000 \\
    \bottomrule
  \end{tabular}
\end{table*}

Tables~\ref{tab:rq3_edge_per_class} and~\ref{tab:rq3_iot_per_class} compare the per-class detection performance of our model and of the baseline \cite{ferrag2024revolutionizing}
on Edge-IIoTset and CICIoT2023, respectively. On Edge-IIoTset (Table~\ref{tab:rq3_edge_per_class}), the two models perform similarly on almost all classes. Both struggle in the class \textit{Fingerprinting} , which is the rarest class in the dataset.
On CICIoT2023 (Table~\ref{tab:rq3_iot_per_class}), the baseline slightly outperform our model on some classes where the highest difference magnitude is on the begnin traffic, about $1.71\%$.
For the binary intrusion detection, both models achieve the same performance on both datasets.

\begin{table*}[ht!]
  \centering
  \footnotesize
  \caption{Per-class test results on CICIoT23 for the SecurityBERT baseline and the proposed model (e128M128, shared, $L{=}1$)}
  \label{tab:rq3_iot_per_class}
  \begin{tabular}{l r cccc cccc}
    \toprule
     & & \multicolumn{4}{c}{SecurityBERT} & \multicolumn{4}{c}{\textbf{Ours}} \\
    \cmidrule(lr){3-6} \cmidrule(lr){7-10}
    Class & Support & F1 & DR & FNR & FPR & F1 & DR & FNR & FPR \\
    \midrule
    BenignTraffic & 3{,}672 & 91.02 & 96.16 & 3.84 & 0.363 & 89.31 & 98.20 & 1.80 & 0.520 \\
    DDoS-ICMP\_Flood & 24{,}192 & 99.99 & 99.98 & 0.02 & 0.001 & 99.84 & 99.94 & 0.06 & 0.047 \\
    DDoS-PSHACK\_Flood & 13{,}859 & 99.96 & 99.96 & 0.04 & 0.004 & 99.92 & 99.88 & 0.12 & 0.004 \\
    DDoS-SYN\_Flood & 13{,}747 & 99.96 & 99.97 & 0.03 & 0.004 & 99.91 & 99.93 & 0.07 & 0.012 \\
    DDoS-RSTFINFlood & 13{,}623 & 99.97 & 99.99 & 0.01 & 0.004 & 99.90 & 99.92 & 0.08 & 0.010 \\
    Recon-PortScan & 279 & 65.75 & 51.61 & 48.39 & 0.010 & 62.22 & 45.16 & 54.84 & 0.000 \\
    DDoS-UDP\_Fragmentation & 964 & 99.84 & 99.69 & 0.31 & 0.000 & 99.38 & 99.07 & 0.93 & 0.002 \\
    DDoS-UDP\_Flood & 18{,}181 & 99.97 & 99.96 & 0.04 & 0.002 & 99.90 & 99.86 & 0.14 & 0.009 \\
    Mirai-greeth\_flood & 3{,}318 & 99.86 & 99.73 & 0.27 & 0.000 & 99.12 & 98.82 & 1.18 & 0.012 \\
    DDoS-TCP\_Flood & 15{,}194 & 99.99 & 99.99 & 0.01 & 0.001 & 99.89 & 99.91 & 0.09 & 0.013 \\
    DNS\_Spoofing & 605 & 69.69 & 55.87 & 44.13 & 0.017 & 69.41 & 60.00 & 40.00 & 0.050 \\
    DoS-TCP\_Flood & 8{,}971 & 99.98 & 99.98 & 0.02 & 0.001 & 99.94 & 99.91 & 0.09 & 0.002 \\
    DoS-UDP\_Flood & 11{,}218 & 99.99 & 99.99 & 0.01 & 0.001 & 99.95 & 99.97 & 0.03 & 0.006 \\
    DDoS-ACK\_Fragmentation & 965 & 99.59 & 99.48 & 0.52 & 0.002 & 99.28 & 99.69 & 0.31 & 0.007 \\
    DDoS-SynonymousIP\_Flood & 12{,}102 & 99.95 & 99.94 & 0.06 & 0.003 & 99.91 & 99.86 & 0.14 & 0.003 \\
    DoS-SYN\_Flood & 6{,}781 & 99.94 & 99.91 & 0.09 & 0.001 & 99.90 & 99.90 & 0.10 & 0.005 \\
    DDoS-ICMP\_Fragmentation & 1{,}534 & 99.48 & 99.67 & 0.33 & 0.007 & 98.82 & 98.57 & 1.43 & 0.009 \\
    Mirai-udpplain & 3{,}025 & 99.95 & 99.97 & 0.03 & 0.001 & 99.01 & 99.67 & 0.33 & 0.033 \\
    Recon-HostDiscovery & 451 & 80.04 & 79.60 & 20.40 & 0.056 & 75.11 & 74.28 & 25.72 & 0.068 \\
    Mirai-greip\_flood & 2{,}543 & 99.88 & 99.92 & 0.08 & 0.003 & 98.60 & 98.03 & 1.97 & 0.014 \\
    MITM-ArpSpoofing & 1{,}053 & 79.07 & 82.34 & 17.66 & 0.175 & 77.06 & 67.81 & 32.19 & 0.055 \\
    Recon-OSScan & 333 & 65.57 & 53.75 & 46.25 & 0.022 & 65.74 & 49.55 & 50.45 & 0.003 \\
    DoS-HTTP\_Flood & 252 & 99.60 & 99.21 & 0.79 & 0.000 & 100.00 & 100.00 & 0.00 & 0.000 \\
    \midrule
    Macro average & - & 93.44 & 92.03 & 7.97 & 0.029 & 92.70 & 90.78 & 9.22 & 0.038 \\
    Benign vs.\ attack & - & 99.77 & 99.64 & 0.36 & 3.840 & 99.72 & 99.48 & 0.52 & 1.797 \\
    \bottomrule
  \end{tabular}
\end{table*}

\begin{table*}[ht!]
\centering
\caption{Comparison between the proposed compressed model and SecurityBERT with PPFLE in terms of model size and inference efficiency. Measurements are reported for a batch size of 10.}
\label{tab:comprehensive_model_comparison}
\small
\setlength{\tabcolsep}{3pt}
\begin{tabular}{lcccccc}
\toprule
\textbf{Model} & \textbf{Params} & \textbf{Size (MB)} & \textbf{HW} & \textbf{Avg. Latency (ms)} & \textbf{Speed/sample (ms)} & \textbf{RAM (MB)} \\
\midrule
\multirow{4}{*}{SecurityBERT} 
& \multirow{4}{*}{11,175,443} 
& \multirow{4}{*}{45} 
& NVIDIA GH100  & $3.10 \pm 10.78$ & 0.31 & 396.00 \\
& & & CPU & $304.27 \pm 7.67$ & 30.43 & 698.00 \\
& & & Simulated Edge  & $810.69 \pm 59.27$ & 81.07 & 66.07 \\
& & & Raspberry Pi 3 & $7459 \pm 404.9$ & 745.97 & 207.29 \\
\midrule
\multirow{4}{*}{\textbf{Ours}} 
& \multirow{4}{*}{1,586,191} 
& \multirow{4}{*}{6} 
& NVIDIA GH100 & $1.46 \pm 0.15$ & 0.15 & 0.0 \\
& & & CPU & $150.03 \pm 8.12$ & 15.00 & 72.00 \\
& & & Simulated Edge & $9.07 \pm 0.56$ & 0.91 & 25.00 \\
& & & Raspberry Pi 3 & $132.22 \pm 30.08$ & 13.42 & 23.80 \\
\bottomrule
\end{tabular}
\end{table*}

Table \ref{tab:comprehensive_model_comparison} reports parameter count, model size, inference latency, per-sample speed, and memory usage on four hardware configurations: NVIDIA GH100, a standard CPU, a simulated edge device, and a Raspberry Pi 3. Across these settings, the proposed architecture consistently exhibits much lower computational cost. The advantage is particularly clear on the simulated edge device, where our model processes a batch in $9.07 \pm 0.56$ ms, corresponding to 0.91 ms per sample, while using only 25 MB of memory. Under the same setting, SecurityBERT requires $810.69 \pm 59.27$ ms per batch and substantially higher memory. On CPU, the proposed architecture also remains about twice as fast while requiring far less memory. These differences are not marginal improvements; they directly affect deployability in constrained IoT environments.

Taken together, the results show that the proposed architecture offers a much more favorable performance, efficiency trade-off than the legacy SecurityBERT-based alternative. While the larger baseline may still achieve slightly better results in some settings, these gains are small relative to the substantial increase in parameter count, storage footprint, inference time, and memory usage. From the perspective of practical IoT deployment, the reduced architecture is therefore not merely a compressed approximation of the baseline, but a more appropriate design choice.

\begin{rqreply}{RQ3}
Among the compared methods, the proposed architecture offers the most favorable trade-off between performance and efficiency for IoT intrusion detection. With PPFLE, it achieves $98.88\%$ accuracy and $94.55\%$ macro F1-score on Edge-IIoTset, outperforming SecurityBERT, while reducing the model size from $45$\,MB to $6$\,MB and the parameter count from $11.2$\,M to $1.6$\,M (about $7\times$ fewer). On CICIoT2023, SecurityBERT remains slightly ahead, but the gap is limited to about $0.7$ percentage points of macro F1-score. This small difference does not justify the much higher inference cost of SecurityBERT: on the simulated edge device, it requires $810.69$\,ms per batch, compared with $9.07$\,ms for our approach.
These results support using the proposed architecture instead of SecurityBERT for intrusion detection in resource-constrained IoT settings.
\end{rqreply}

\subsection{Discussion}

The results of the three research questions give a coherent view of the strengths and limitations of the proposed architecture for IoT intrusion detection. Overall, they show that the model can be strongly compressed without a proportional loss in detection performance.

Firstly, moderate compression preserves detection performance, whereas the most aggressive compression has a measurable cost. The $e128M128$ configuration provides the best trade-off: it reaches $98.06\%$ macro F1-score, close to the best configuration ($e256M64$), which requires about $3.4\times$ more core parameters and $1.5\times$ more compute. The most compact configuration, $e64M64$, reduces compute by a factor of $3.2$ but loses $2.4$ points of macro F1-score and is less stable across seeds. All configurations reached an accuracy above $98\%$, and only the macro F1-score revealed these differences. Because intrusion detection datasets are highly imbalanced, models should be evaluated with class-sensitive metrics such as the macro F1-score, not with accuracy alone. We hypothesize that the latent bottleneck acts as a filter that retains the information relevant to traffic classification; however, our experiments do not test this hypothesis directly.

Secondly, The two strategies (stacking and sharing) reach comparable macro F1-scores (differences of at most $0.33$ points) and have identical compute and peak training memory at a given depth. Their advantage is therefore not predictive: sharing keeps the number of non-embedding parameters constant ($306$k), whereas stacking adds about $198$k parameters per layer. This reduces the size of the stored model by $39\%$ in $L{=}2$ and $66\%$ in $L{=}4$, which is important in IoT and edge devices with limited storage. However, additional depth brought little benefits with either strategy. In our setting, a single layer is thus sufficient, and parameter sharing is the preferable strategy only when more depth is needed.

Finally, the proposed architecture remains competitive with a much larger Transformer model. With the same privacy-preserving encoding (PPFLE), it outperforms SecurityBERT on Edge-IIoTset when using the word level tokenization instead of the Byte level tokenization and is only $0.7$ points behind on CICIoT2023. At the same time, it has about $7\times$ fewer parameters ($1.6$\,M vs.\ $11.2$\,M) and is about $89\times$ faster on the simulated edge device. This suggests that, for IoT intrusion detection, increasing model size beyond a certain point brings only marginal gains at a high memory and latency cost. In constrained environments, model selection should therefore be based on the performance/efficiency trade-off rather than on predictive performance alone.
However, the study is subject to some limitations. First, all models were trained for at most 10 epochs; the deeper configurations ($L{=}4$) had not converged, so their performance may be underestimated. Second, some classes remain difficult to detect regardless of the configuration (\textit{Fingerprinting}, \textit{Ransomware} and \textit{Backdoor} on Edge-IIoTset; \textit{Recon-OSScan}, \textit{Recon-PortScan} and \textit{DNS Spoofing} on CICIoT2023), and neither compression choices nor additional depth resolved them. Third, inference latency was measured on a simulated edge device and should be confirmed on physical hardware. Finally, the evaluation is limited to two IoT datasets, so the results may not generalize to broader network environments.

\section{Conclusion}\label{sec:conclusion}

This work proposed a parameter-efficient language model for IoT intrusion detection. The retained configuration ($e128M128$) uses about $1.6$\,M parameters in total, of which $306$k are non-embedding, for a model size of about $6~$MB.
With the same privacy-preserving encoding (PPFLE), it outperforms SecurityBERT ($11.2$\,M parameters, $45$\,MB) by about $2$ points of macro F1-score on Edge-IIoTset and is within $0.7$ points on CICIoT2023. On a simulated edge device, it requires $9.07$\,ms per batch and $25$\,MB of memory, i.e., it is about $89\times$ faster than SecurityBERT ($810.69$\,ms). A single Transformer layer is sufficient in our setting. When more depth is needed, parameter sharing matches stacking with up to $66\%$ fewer parameters. These results make the proposed model well suited to resource-constrained IoT environments. Future work will address the classes that remain difficult to detect, validate the model on physical edge hardware, and evaluate it on additional datasets.
\section*{Acknowledgements}

This work was carried out with the support of the ROMEO Regional Computing Center. The authors also gratefully acknowledge Saint Jean Ingénieur, a Cameroonian private institution of higher education, for their generous funding and support, which made this research possible.

\bibliographystyle{plainnat} 
\bibliography{references}

\end{document}